\documentclass[a4paper,12pt]{article}
\usepackage[cp1251]{inputenc}
\usepackage{graphics}
\usepackage{amssymb,amsmath,latexsym}
\usepackage[english]{babel}
\usepackage{indentfirst}
\usepackage{graphicx,times}
\usepackage{natbib}
\begin{document}

\centerline {\bf Can Rossby waves explain the cyclic magnetic}
\centerline {\bf activity of the Sun and solar-type stars? }

\bigskip
\centerline {E.A. Bruevich $^{a}$ , I. K. Rozgacheva $^{b}$}
\bigskip
\centerline  {{$^a${Sternberg Astronomical Institute, MSU, Moscow,
Russia }}}

\centerline {$^b${VINITI RAS, Moscow, Russia}}
\bigskip
\centerline {E-mail: $^a${red-field@yandex.ru},
$^b${rozgacheva@yandex.ru}}

\bigskip

{\bf ABSTRACT}

Magnetic activity is a global property of the Sun; the complex
processes of solar activity are connected with the solar magnetic
fields. For solar-type stars and the Sun magnetic activity depends
on the physical parameters of the star. In this article we study the
relationships between the duration of activity cycle and effective
temperature for solar-type stars and the Sun. We've tried to explain
these relationships due to the existence of layer of laminar
convection near the bottom of the convective zone of the star. In
this layer the Rossby waves are formed. They generate the primary
poloidal magnetic field, which is the source of energy of the
complex phenomena of magnetic activity.

KEY WORDS: Sun; Magnetic Activity; Solar-Type Stars; Activity
Cycles; Convection, Rossby Waves.

\section{Introduction}

Magnetic activity of the Sun is called the complex of
electromagnetic and hydrodynamic processes in the solar atmosphere.
The analysis of active regions (plages and spots in the photosphere,
flocculae in the chromosphere and prominences in the corona of the
Sun) is required to study the magnetic field of the Sun and the
physics of magnetic activity. The very important property of solar
irradiance is its cyclical nature.

J. Hale and S. Nicholson discovered the solar magnetic field. After
systematic studies they stated the magnetic law of polarity of the
spots (Hale-Nicholson law) and the changes of their characteristics
during the solar cycle. They found that the periodic variation of
the number of spots according to a 11 years cycle constitutes half
of the 22 years cycle of the evolution of the solar magnetic field.
The most outstanding aspect is the inversion of East-West polarity
of the magnetic fields of the spots of the active areas which
accompanies the 11 years cycle [1].

J. Hale and S. Nicholson found that within one solar cycle in
bipolar magnetic fields all the spots p from one hemisphere and all
the spots f from a different hemisphere have the same polarity, and
in the next cycle the polarities of all these spots are changed to
the opposite. Change in such a way that each solar cycle is the
epoch of the constant polarity of heliomagnetic fields: the change
of cycles corresponds to the change of its polarity, and the total
magnetic cycle contains two adjacent spot's cycles. These laws Hale
- Nicholson show, on the one hand, that the mechanism of generation
of heliomagnetic field affects the oscillatory manner, producing a
fairly regular (quasi periodic) reverse generated field. On the
other hand, these laws can be seen that the generation as
heliomagnetic field and fluctuations of solar activity is performed
with one and the same mechanism. Thus, it becomes clear that the
theory of the solar cycle is a global problem of magnetic
hydrodynamics of the Sun.

The differential rotation of the Sun is accompanied by energy costs
for overcoming the forces of viscosity (primarily turbulent
viscosity created by small-scale convective motions in granules and
supergranuls). The angular velocity of rotation at different
heliographic latitudes may be equalized during a few turns of the
sun without any supporting mechanism. According to modern concepts
[2] such a supporting mechanism is the mechanism of meridional and
radial momentum transfers in the convective zone of the Sun with the
help of giant cells (which are influenced by the rotation of the
Sun). These giant cells form the spiral macro turbulence, in which
the velocity vortex is not orthogonal to the velocity. Giant
convection cells were detected by magnetic fields observations: the
fields with opposite polarities of magnetic field alternate (for
different longitudes) with prevailing wave number m = 6. They,
perhaps, are the manifestation of Rossby waves in the convective
layer of the Sun.

The solar cycle (or solar magnetic activity cycle) is the periodic
change in the sun's activity (including changes in the levels of
solar radiation and ejection of solar material) and appearance
(visible in changes in the number of sunspots, flares, and other
visible manifestations). Solar cycles have an average duration of
about 11 years. For hundreds of years of solar activity observations
the duration of this cycle (called the Schwabe cycle) vary from 9 to
12 years. Differential rotation of the sun's convection zone
consolidates magnetic flux tubes, increases their magnetic field
strength and makes them buoyant. For the first time such a model
considered by G. Babcok [3]. As they rise through the solar
atmosphere they partially block the convective flow of energy,
cooling their region of the photosphere, causing 'sunspots'. The
Sun's apparent surface, the photosphere, radiates more actively when
there are more sunspots. Satellite monitoring of solar luminosity
since 1980 has shown there is a direct relationship between the
solar activity (sunspot) cycle and luminosity with solar cycle
peak-to-peak amplitude of about 0.1\% [1].

At the present time in solar and stellar physics also study multiple
and changing cycles with relatively small-amplitude: quasi-biennial,
semi centennial and century activity cycles [4 - 7]. In a simple
model of a stochastically excited solar dynamo [8] it was shown that
the modern observations of the Sun have established the profiles of
the flows in the solar interior. It was found that the differential
rotation of the sun has a strong shear in a thin layer at the base
of the convective region, called the tachiocline. This layer
comprises less than 3\% of the solar radius. The strong magnetic
fields observed in the solar surface must be stored in this deep
layer of the sun for a timescale of the order of the solar cycle. In
this region, the magnetic flux tubes must be intense enough as to
travel through the convection zone without being destroyed by the
turbulent movements. In the modern models of heliomagnetic dynamo it
is believed that the toroidal field is created from the poloidal by
differential rotation of the convective zone of the Sun as was first
considered at [3]. In his model after many rotations, the field
lines become highly twisted and bundled, increasing their intensity,
and the resulting buoyancy lifts the bundle to the solar surface,
forming a bipolar field that appears as two spots, being kinks in
the field lines. The leading spot of the bipolar field has the same
polarity as the solar hemisphere, and the trailing spot is of
opposite polarity. The leading spot of the bipolar field tends to
migrate towards the equator, while the trailing spot of opposite
polarity migrates towards the solar pole of the respective
hemisphere with a resultant reduction of the solar dipole moment.
This process of sunspot formation and migration continues until the
solar dipole field reverses (after about 11 years).

Thus the solar magnetic fields are created by dynamo processes. In
them the poloidal and toroidal fields are changed to each other; in
$\alpha$-effect the poloidal field lines are stretched out and wound
around the Sun by differential rotation forming toroidal field
lines; twisting of the toroidal field lines into poloidal filed
lines is caused by effects of solar rotation, so called
$\Omega$-effect.

The $\alpha\Omega$ - dynamo theory which is based on the hypothesis
about the generation of the magnetic field due to the differential
rotation of the Sun in the turbulent convective shell can describe
the main features of solar magnetic activity [5, 6]. Solar cycle
models based on what is now called the Babcock-Leighton mechanism:
1) generation of toroidal field from the poloidal field due to the
differential rotation of the convective shell ($\Omega$- effect); 2)
generation of poloidal field from bipolar magnetic regions of
toroidal field due to the differential rotation and the turbulent
viscosity ($\alpha$- effect). The $\alpha\Omega$ - dynamo theory
which is based on the hypothesis about the generation of the
magnetic field due to the differential rotation of the Sun in the
turbulent convective shell can describe the main features of solar
magnetic activity [7, 9]. This theory is simulates well the
following phenomena of magnetic activity such as the formation of
strong local bipolar magnetic fields (0,1 Tesla), the cyclicity of
magnetic activity and Sporer's law. According to the theory of
$\alpha\Omega$ - dynamo the faster the star rotates, the higher the
value of parameter of differential rotation (the difference between
the periods of rotation of the polar regions and of the equator) and
also the magnetic activity is higher. However, were found that the
observations do not confirm this conclusion [10].

We use the HK Project data [10, 11]. These observations are the
longest-running program to monitor stellar activity cycles similar
to the 11-year sunspot cycle. Almost 100 stars have been observed
continuously since 1966; at present the project is monitoring
long-term changes in chromospheric activity for approximately 400
dwarf and giant stars. The HK-Project uses a specially-designed
instrument to measure the amount of light from active magnetic
regions in stars. This light comes from calcium atoms that have lost
one electron each. The different wavelengths of light emitted by
these atoms were labeled long ago. The $H$ and $K CaII$  light gave
this project its name. This light comes from the upper levels of the
Sun near active magnetic regions that we can see, like sunspots.
Other stars are too far away to see these features on their
surfaces. Studying the relative strength of these two wavelengths of
calcium light from distant stars similar to our Sun gives an
indirect measure of the amount of surface activity on the stars -
"starspots". Using this method, astronomers have been able to follow
cycles similar to the sunspot cycle that has been observed on the
Sun for centuries. For solar-type stars from HK-project [10, 11]
were found that the differential rotation increases with the growth
of the average period of rotation. This feature of the rotation of
stars creates problems for the theory of $\alpha\Omega$ - dynamo
[12].

Complex relationship of poloidal and toroidal magnetic fields
indicates that in the convective shell of the Sun there are not less
than two working mechanisms of the solar dynamo. They have a variety
of spatial scales and characteristic times of their formation. In
the framework of the hypothesis about the only one turbulent
convective shell, it is impossible to ensure sustainable separation
of small-scale and large-scale hydromagnetic dynamo.

In this paper we study the relationships between the duration of
activity cycle $T_{cyc}$ and effective temperature $T_{eff}$ for
solar-type stars and the Sun. We explain these relationships due to
the existence of Rossby waves which are formed at the bottom of the
convective zone of the stars and the Sun. The Rossby waves conserve
vorticity and owe its existence to the variation of the Coriolis
force with latitude. The Rossby waves are connected with the primary
poloidal magnetic field of a star, which is the source of energy of
the complex phenomena of magnetic activity.

The main our assumption is: the time of generation of Rossby waves
$t_g$ corresponds to duration of the activity cycle $T_{cyc}$. We
show that theoretical dependence of the time of generation of Rossby
waves $t_g$ versus $T_{eff}$ (the basic parameter of a star)
describes well the connection between the star's duration of the
activity cycle $T_{cyc}$(obtained from observations of solar-type
stars and the Sun) and their $T_{eff}$.

\section{"The rotation period - the effective temperature"
and "the duration of the cycle of activity - the effective
temperature" relationships}

To refine the theory of the solar dynamo search of relationships
between the physical parameters of solar-type stars is necessary.
These are the effective temperature, the rotation period, the
duration of cycles, the magnetic field and the age.

In the Table 1 we present the information about the periods of
rotation $P_{rot}$ and effective temperatures of stars $T_{eff}$,
about the duration of their 11-year $T_{11}$ and quasi-biennial
$T_{2}$ cycles.

The periods of cycles of activity are given according to the
calculations of cyclic periodicities of HK-project stars by the
authors of this work with help of Fourier analysis of light curves
of stars [5]. We also used the earlier definition of "11-year"
periods by the authors of HK-project ($T^{HK}_{11}$) [11] for a
sample of 52 stars and the Sun.

We used the data of Table 1 and found the statistically significant
relationships "the rotation period - the effective temperature" and
"the duration of activity cycle - the effective temperature", see
Fig. 1- 3.

\begin{table}
\caption{Results of our calculations of $T_{11}$ and $T_{2}$ values
and HK-project $T^{HK}_{11}$ calculations for 52 stars and the Sun}
\vskip12pt
\begin{tabular}{clclclclclclclcl}

\hline
    1&    2    &     3   &          4    &     5     &     6     &     7     &   8     \\ \hline
  No & Star    &Spectral &$P_{rot}, days$& $T_{eff}$, K &$T^{HK}_{11}$, &$T_{11}$,   &$T_{2}$ \\
     & on the HD& class   &(Soon et         &(Allen        &   years       &years       &years  \\
     & catalog &         &  al 1996)       & 1977)        &               &           &         \\ \hline
   1 &  Sun   & G2-G4   &    25           &     5780     &  10,0        &  10,7     &   2,7     \\
   2 &HD1835  & G2,5    &     8           &     5750     & 9,1          &   9,5     &   3,2     \\
   3 &HD3229  & F2      &     4           &     7000     & 4,1          &   -     &   -     \\
   4 &HD3651  & K0      &     45           &    4900     & 13,8          &   -     &   -     \\
   5 &HD4628  & K4      &    38,5         &     4500     & 8,37          &   -     &   -     \\
   6 &HD20630  & G5     &     9,24        &     5520     & 10,2          &   -     &   -     \\
   7 &HD26913 & G0      &     7,15        &     6030     & 7,8           &   -     &   -     \\
   8 &HD26965 & K1      &     43          &     4850     & 10,1          &   -     &   -     \\
   9 &HD32147 & K5      &     48          &     4130     &12,1           &   -     &   -     \\
   10&HD10476 & K1      &    35           &     5000     &  9,6         & 10       &    2,8    \\
   11&HD13421 & G0      &    17           &     5920     & -            & 10       &    -      \\
   12&HD18256 & F6      &    3            &     6450     & 6,8          & 6,7      &  3,2     \\
   13&HD25998  & F7     &    5            &     6320     &  -           &  7,1     &   -       \\
   14&HD35296 & F8      &   10            &     6200     &  -           & 10,8      &   -       \\
   15&HD39587 & G0      &    14           &     5920     &  -           & 10,5      &    -      \\
   16&HD75332 & F7      &    11           &     6320     &  -           &  9        &   2,4    \\
   17&HD76151 & G3      &    15           &     5700     &  -           &       -   &    $2,52^*$  \\
   18&HD76572 & F6      &    4            &     6450     &   7,1        & 8,5      &    -     \\
   19&HD78366 & G0      &    9,67         &     6030     &              &  10,2    &   -     \\
   20&HD81809  & G2     &   41           &      5780    & 8,2          & 8,5       &   2,0      \\
   21&HD82885 & G8      &   18           &      5490    & 7,9          & 8,6       &   -       \\
   22&HD100180& F7     &   14            &      6320    & 12           & 8         &    -       \\
   23&HD103095 & G8     &   31           &      5490    & 7,3          & 8         &    -       \\
   24&HD114710 & F9,5   &    12          &      6000    & 14,5        & 11,5       &   2      \\
   25&HD115383 & G0     &   12           &      5920    &  -           & 10,3      &    3,5      \\
   26&HD115404 & K1     &    18          &      5000    & 12,4         & 11,8      &   2,7     \\
   27&HD120136 & F7     &   4            &      6320    & 11,6         & 11,3       &    3,3      \\
   28&HD124570& F6      &    26          &      6450    & -            & -         &   2,7     \\
   29&HD129333 & G0     &   13           &      5920    & -            & 9         &    3,2      \\
   30&HD131156 & G2     &    6           &      5780    & -            & 8,5       &   3,8     \\

\hline
\end{tabular}
\end{table}

\begin{table}
\caption{Table 1 - continued}

\vskip12pt
\begin{tabular}{clclclclclclclcl}

\hline
    1&    2    &     3   &          4    &     5     &     6     &     7     &   8     \\ \hline
  No & Star    &Spectral &$P_{rot}, days$& $T_{eff}$, K &$T^{HK}_{11}$, &$T_{11}$,   &$T_{2}$ \\
     & on the HD& class   &(Soon et         &(Allen        &   years       &years       &years  \\
     & catalog &         &  al 1996)       & 1977)        &               &           &         \\ \hline
    31&HD143761 & G0     &   17            &     5920    & -            &   8,8     &    -      \\
   32&HD149661 & K2     &    21          &      4780    & 14,4         & 11,5      &   3,5     \\
   33&HD152391 & G7   &     11           &      5500    & 10,7         &  -        &     -     \\
   34&HD154417 & F8   &     7,8           &     6100    & 7,4          &   -        &    -     \\
   35 &HD155875 & K1    &     30           &      4850    & 5,7          &   -        &    -     \\
   36 &HD156026 & K5    &    21          &       4130    & -        &  11        &    -     \\
   37 &HD157856 & F6     &    4           &       6450   & -            & 10,9     &   2,6     \\
   38 &HD158614 & G9     &   34           &       5300   & -            & 12       &    2,6      \\
   39 &HD160346 & K3     &    37          &    4590      & 7           & 8,1        &   2,3     \\
   40 &HD166620 & K2      &   42           &       4780   & 15,8          & 13,7        &     -       \\
   41 &HD182572 & G8     &    41          &    5490      & -            & 10,5       &   3,1     \\
   42 &HD185144 & K0     &   27           &       5240   & -             & 8,5       &    2,6     \\
   43 &HD187681 & F8     &   10           &      6100    & 7,4            &  -        &    -     \\
   44 &HD188512 & G8     &   17           &       5490   & -            &  -        &    4,1     \\
   45 &HD190007 & K4     &    29          &      4500     & 10         & 11         &   2,5     \\
   46 &HD190406 & G1     &    14          &     5900     & 8        &   -        &     -     \\
   47 &HD201091 & K5     &   35           &      4410     &          & 13,1      &    3,6      \\
   48 &HD201092 & K7     &    38          &      4160     &          & 11,7      &   2,5     \\
   49 &HD203387 & G8     &   30            &       5490    & -            &  -        &    2,6      \\
   50 &HD206860 & G0     &    9          &     6300      & 6,2         & -        &  -     \\
   51 &HD216385 & F7     &    7           &       6320     & -           &  7        &   2,4    \\
   52 &HD219834 & K2     &    43          &      4780     & 10         & 11         &   2,5     \\
   53 &HD224930 & G3     &    33          &     5750    & 10,2        & -        &   -     \\
\hline
\end{tabular}

\end{table}

Linear regression equation has the following form:
$$ \log P_{rot}=15,7 -  3,87 \cdot \log T_{eff}   \eqno (1)$$

The linear correlation coefficient (Pearson's correlation
coefficient) in the regression equation (1) is equal to 0,73.
According to Pearson's cumulative statistic test (which
asymptotically approaches a $ \chi^2$ - distribution) the linear
correlation between the $P_{rot}$ and $T_{eff}$  is statistically
significant at a 0,05 level of significance.

Thus our data set of the rotation periods and the effective
temperatures of stars shows the    following power-law dependence
$$ P_{rot} \sim T_{eff}~^{-3,9}  \eqno (2) $$

This ratio follows from the law of radiation Stefan - Boltzmann,
from the well known "mass-luminosity" relationship and from the
conservation of the angular momentum of the stars [7].

\begin{figure}[h!]
\includegraphics[width=140mm]{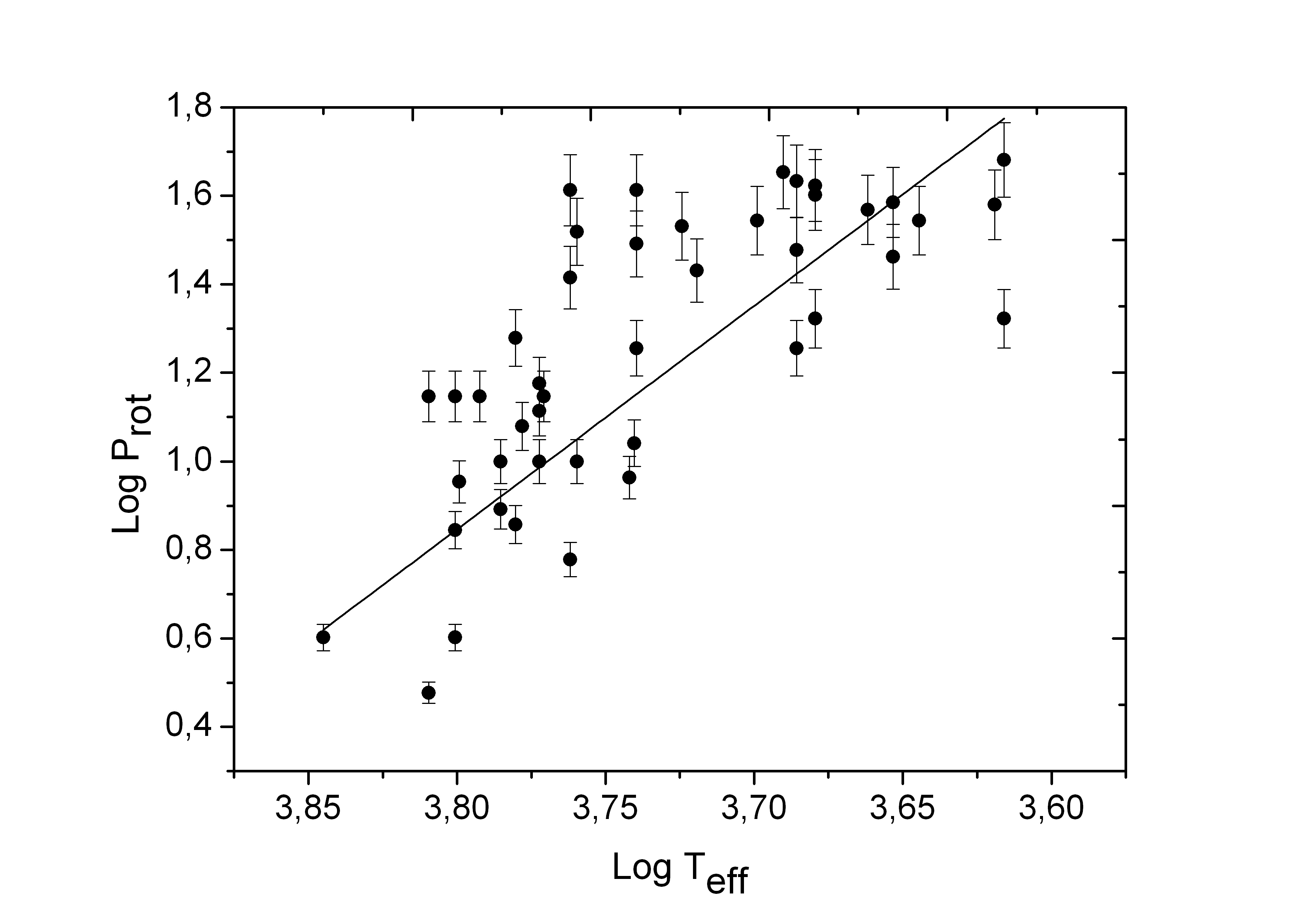}
\caption{Diagram "the rotation period $P_{rot}$ - the effective
temperature $T_{eff}$". The line shows the linear regression on a
data set. The error bars corresponds to $1\sigma$ scatter of linear
regression} \label{Fi:Fig1}
\end{figure}
\medskip

The linear regression equation for points of diagram at the Fig. 2
is of the form:

$$ \log T_{11}= 5,15 -  1,11 \cdot \log T_{eff}   \eqno (3)$$

The linear correlation coefficient (Pearson's correlation
coefficient) in the regression equation (3) is equal to (- 0,67).
According to Pearson's cumulative statistic test (which
asymptotically approaches a
 $ \chi^2$ - distribution)
the linear correlation between the $T_{11}$  and $T_{eff}$ is
statistically significant at a 0,05 level of significance.

Thus, for the investigated sample of stars the periods of "11-year"
cycles $T_{11}$   and their effective temperatures $T_{eff}$  are
connected in power-law dependence:
$$ T_{11} \sim T_{eff}~^{-1,1}  \eqno (4) $$

\begin{figure}[h!]
\includegraphics[width=140mm]{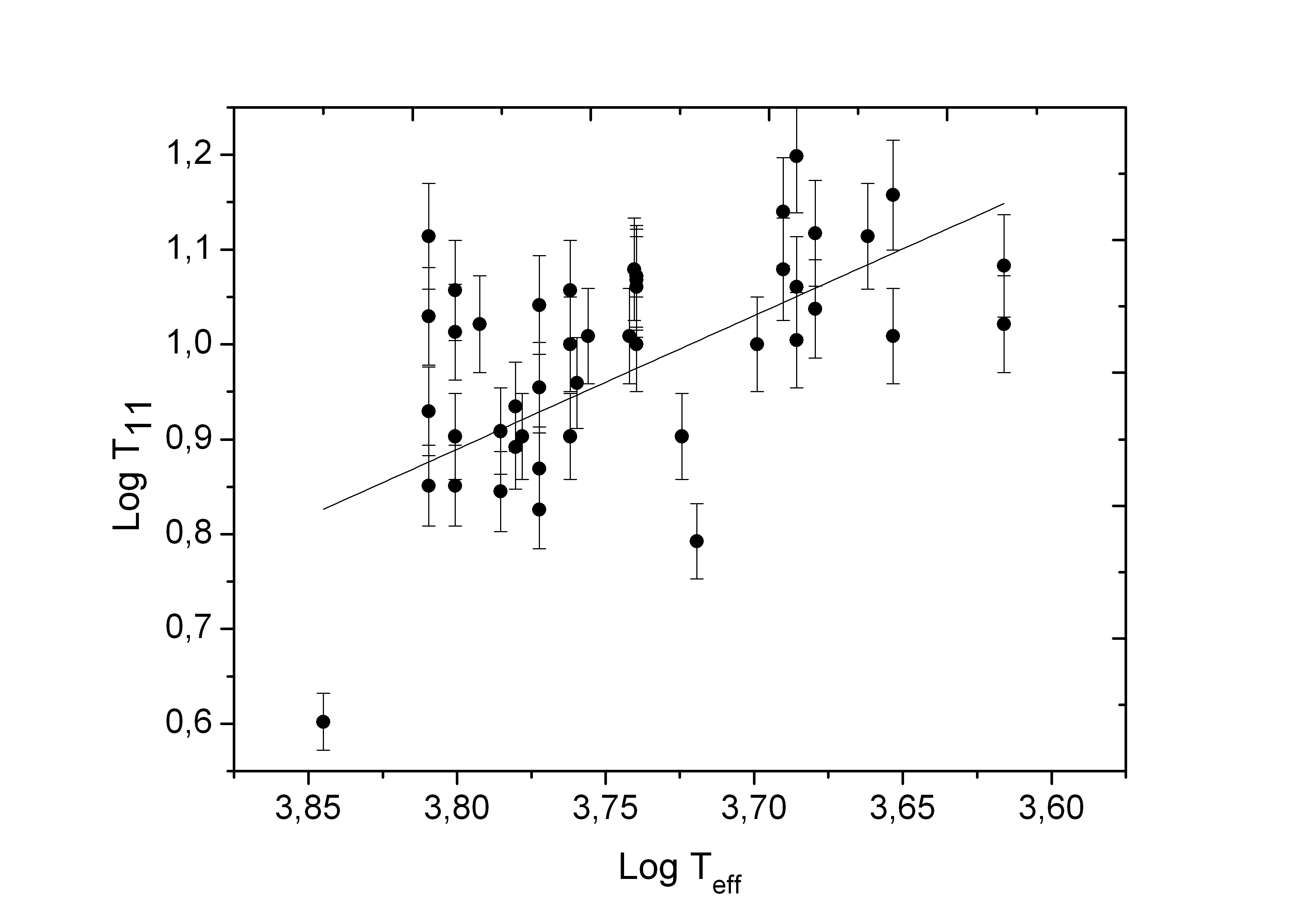}
\caption{Diagram "the duration of the cycle $T_{11}$ - the effective
temperature $T_{eff}$". The line shows the linear regression on a
data set. The error bars corresponds to $1\sigma$ scatter of linear
regression} \label{Fi:Fig2}
\end{figure}
\medskip

We used the data of observations of variations of chromospheric
radiation of solar-type stars and the Sun from the Table for
statistical analysis and the search if there is a possible linear
relationship between the duration of the quasi-biennial cycles of
the stars  $T_2$ and their effective temperatures $T_{eff}$.

The diagram of  "the duration of the cycle - the effective
temperature" ($T_{2}$ is the period of quasi-biennial cycles) for
the Sun and stars from the Table is shown in Fig. 3.

Linear regression equation for points of diagram at the Fig. 3 is of
the form:
$$ \log T_2 = 3,46 -  0,79 \cdot \log T_{eff}   \eqno (5)$$

The linear correlation coefficient (Pearson's correlation
coefficient) in the regression equation (4) is equal to  (- 0,51).
According to Pearson's cumulative statistic test (which
asymptotically approaches a $ \chi^2$ - distribution ) the linear
correlation between the $T_2$  and $T_{eff}$  is statistically
significant at a 0,1 level of significance.

Thus, for the investigated sample of stars their periods of
quasi-biennial cycles $T_2$  and their effective temperatures
$T_{eff}$ are connected as power-law dependence:

$$ T_2 \sim T_{eff}~^{-0,79}  \eqno (6) $$

\begin{figure}[h!]
\includegraphics[width=140mm]{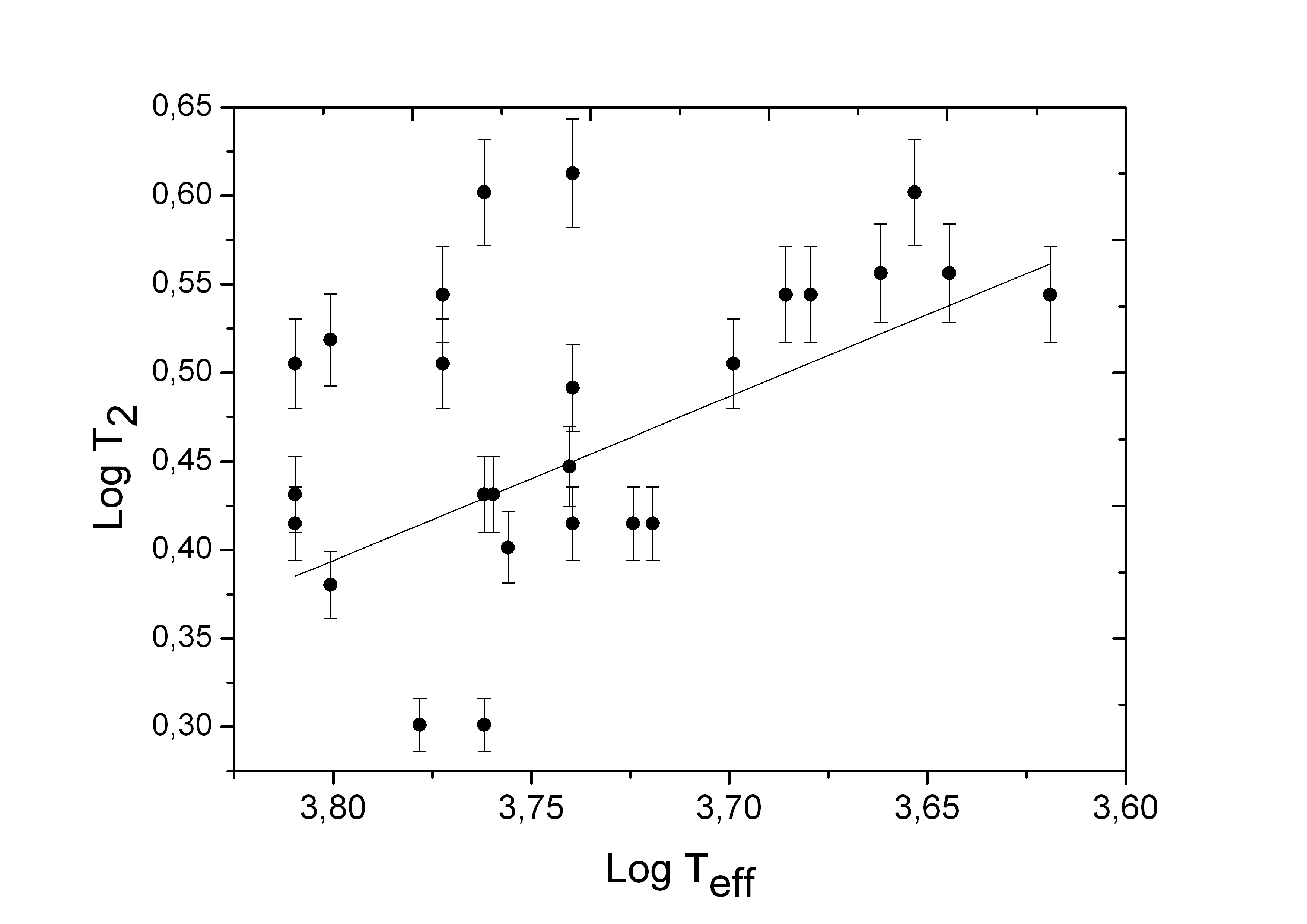}
\caption{Diagram "the duration of the cycle $T_{2}$ - the effective
temperature $T_{eff}$". The line shows the linear regression on a
data set. The error bars corresponds to $1\sigma$ scatter of linear
regression} \label{Fi:Fig3}
\end{figure}
\medskip

A numerical MHD simulation of hydro magnetic dynamo in case of the
fully turbulent, the fully laminar convective shells and also for
the convective shell which consists of two layers (the turbulent and
the laminar) was made in [13]. We have taken into account these
results of numerical MHD simulation for our model estimations of
solar magnetic cycle's duration.

\section{The model of solar magnetic fields generation}

These properties of the global magnetic activity of the Sun and the
exponential dependencies of Fig. 2, 3 can be understood in the
framework of the scheme of generation of magnetic fields: at the
bottom of convective zone that are heated plasma with help of
photons from radiation zones the giant convection cells are formed.
There are conditions for convective transfer from the account of
only radiant plasma viscosity [6]. The rising element of volume
creates a pressure gradient in plasma. The direction depends on the
direction of the velocity of an element, the Coriolis acceleration
and differential rotation. The pressure gradient will be spread in
the spherical shell along the lines of latitudes as Rossby wave.

P. Gilman [14] considered the model of heliomagnetic dynamo, in
which Rossby waves that formed in the convective zone of the Sun
serve the elements of helical turbulence due to the presence of
significant temperature gradient. The vertical movements of the
Rossby waves create the toroidal magnetic field from large-scale
vertical magnetic fields. Rossby waves transfer them to the poles
thus creating the poloidal field.  A new toroidal field of opposite
sign is formed with help of this poloidal field. In this simplified
Gilman's model the quasiperiodic reverse of heliomagnetic fields
were obtained (with the highest distinction from the observed solar
cycle period - instead of 22 years is equal to about 2 years). This
disagreement is probably caused by the fact that in the Sun there
are not usual Rossby waves that are formed due to the presence of
the latitudinal temperature gradient, but there are the giant
convection cells which are rather twisted by the Coriolis force,
formed due to vertical temperature gradient in the convective zone
of the Sun. Later P. Gilman [15] has modified this model with the
use of Rossby waves in the theory of hydromagnetic dynamo. It was
also analyzed in detail the dynamo action from a typical Rossby wave
motion and compared it with the solar cycle. But the quasiperiodic
reverse of heliomagnetic fields by new Gilman's model was also
obtained as equal to 2 years.

We also took into consideration the main assumptions which were
postulated in [15]. We know that around the base of convective shell
the plasma temperature reaches $T\approx 2\cdot 10^6~K$. Plasma is
fully ionized at such temperatures. Interaction of radiation with
plasma is carried out by the scattering of photons by electrons; if
the characteristic energy of photons does not exceed $kT$ (where $k$
is the Boltzmann constant). The motion of the plasma in the Rossby
wave is slowed by the radiant viscosity of plasma. This viscosity
inhibits the directed motion of electrons (the characteristic time
is a fraction of a second), faster than motion of protons.
Therefore, in the Rossby wave the electric current appears. This
current creates a poloidal field. Lines of force of this poloidal
field have the wave-like structure. This field is concentrated
around the base of convective envelope in the equatorial and
medium-sized heliolatitudes. In the polar heliolatitudes lines of
force of the poloidal field come to the atmosphere of the Sun, see
Fig. 4.

\begin{figure}[h!]
 \centerline{\includegraphics[width=90mm]{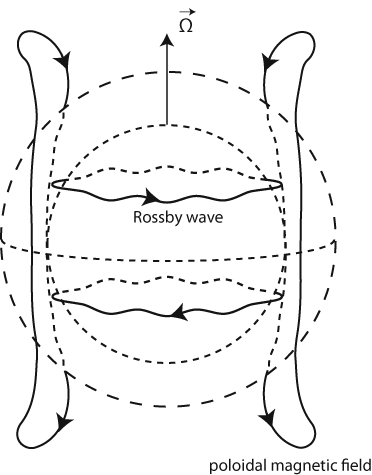}}
\caption{The scheme of the Rossby waves and the poloidal magnetic
field} \label{Fi:Fig4}
\end{figure}

On the length of the latitude the whole number of Rossby waves is
packed. Therefore, the length of these waves should be of the order
of  $\lambda \approx (1/m) 2 \pi R_I \cos \phi $  where $\phi$ is
the latitude of parallels, along which the wave applies, $R_I\approx
0,7 R_{Sun}$   is the radius of the base of the convective envelope,
m is the integer number of  Rossby waves. For giant cells which are
observed in photosphere  the wave number is of order $m=6$, see [2].
At different latitudes the various Rossby waves and different
poloidal magnetic fields are generated. The ordered geometric
structure of the spicules indicates the regularity of poloidal
field. Therefore, the lines of force of poloidal fields, created by
Rossby waves, must be regular and not be entangled due to the
large-scale convection. So the convection around the base of
convective envelope must be laminar. The length of the Rossby wave
is approximately equal to the thickness of the shell of laminar
convection.

The characteristic time t of the convective ascent of the plasma
element which is heated by photons from the radiation zone as was
shown in [6] is equal to:
$$ t \approx \frac {\nu}{gh \left( \frac
{1}{\rho} \frac {\partial \rho}{\partial T} \right) \Delta T} \sim
const $$

where $\nu$  is the coefficient of viscosity of the plasma, $g$ is
the free fall acceleration, $h$ is the thickness of the shell of
laminar convection,  $ \left( \frac {1}{\rho} \frac {\partial
\rho}{\partial T} \right) $  is coefficient of thermal expansion of
the plasma, $ \Delta T$  is the gradient in the layer.

The average Archimedean acceleration of a plasma element is
approximately equal to $$ a\approx g  \left( \frac {1}{\rho} \frac
{\partial \rho}{\partial T} \right) \Delta T $$

 Then the average
Archimedean speed of a plasma element is $ V \approx at \approx
\frac{\nu}{h}$.

The Coriolis forces turn and stretch the convective cell along a
parallel, so that the length of the path of element of the plasma is
comparable with the length of parallels $l=2 \pi \cos \varphi$.

Average acceleration of the Coriolis force on this Parallels is in
the order of magnitude is equal to $a_c= 2V\Omega_{Sun}\sin \varphi
\approx 2 \frac {\nu}{h}\Omega_{Sun}\sin \varphi$  , where $
\Omega_{Sun}$  is the angular velocity of rotation of the Sun around
the base of convective shell.

The time interval which is necessary for the generation of Rossby
waves and poloidal magnetic field on the order of magnitude is equal
to:
$$t_g \approx \sqrt{\frac{2l}{a_c}} \approx \sqrt { \frac { \pi R_I \cos \varphi}{\frac{\nu}{h}
\Omega_{Sun} \sin \varphi}}  \eqno (7) $$

We assumed that the time of generation of Rossby waves $t_g$  is the
characteristic time parameter of solar magnetic dynamo $t_g$ and
corresponds to duration of the activity cycle $T_{cyc}$.

The generation time increases with decreasing of helio-latitude.
Therefore, the first waves appear at high latitudes. We'll consider
that the poloidal field, connected with Rossby waves have to be the
primary for the magnetic activity. $\Omega$ - effect generates a
toroidal field with help of the primary poloidal in average
turbulent layer of the convective shell. Magnetic loops of toroidal
fields are generated because of the turbulence. Rossby wave
compresses these loops and stimulates the formation of spots. So
Rossby waves lead to the formation of "active latitudes". The
meridional fluxes around the base of convective zone are directed
from the poles to the equator [16]. They move Rossby waves closer to
the equator. The consequence of this will be the slope of the lines
of force of poloidal field in the direction to the equator (the axis
of symmetry of the field turns). A toroidal field moves too, so in
the photosphere there is Sp?rer's law. The $\alpha$ - effect
generates the poloidal field from the toroidal field with the
opposite polarity in relation to the primary poloidal field.

The interaction of the new poloidal field with a primary poloidal
field stimulates the polar magnetic activity, hat weakens these
fields. Therefore the new a toroidal field, which appears from the
weakened poloidal field, will be weaker than the previous one
toroidal field.

The local magnetic activity, which the new a toroidal field will
cause in the second half of the magnetic cycle also will be weaker.
A new a toroidal field is directed opposite to the direction of
Rossby wave propagation. It weakens the electric currents that exist
in these waves.

This process leads to a weakening of the primary poloidal field. At
the end of the previous stage of the magnetic cycle the  $\alpha$ -
effect generates the poloidal field. Its direction is close to the
direction of primary poloidal field; if the latter does not
experienced significant changes.

In this case, the primary poloidal will increase. From the primary
poloidal the $\Omega$- effect generates a toroidal field, which is
stronger than the previous toroidal field. Stronger field stimulates
stronger activity. So the Gnevyshev-Ol's rule is explained by the
contribution of the primary poloidal field to the evolution of the
toroidal field.

If the primary poloidal field has experienced a significant turn,
the rule Gnevyshev-Ol's rule is violated as it was observed for the
cycles 22 and 23. The energy of the Rossby waves is the source of
magnetic activity. Therefore, these waves gradually decrease. The
beginning of a new cycle can be delayed until the formation of the
new Rossby waves. This explains the changes the durations of the
cycles. If the generation of Rossby waves will slow down in the
mid-latitudes due to the penetration of turbulent motions in the
lower laminar layer, it will accelerate the meridional transfer of
momentum, which will speed up the rotation of the mid-latitudes.

The time of generation of Rossby waves $t_g$ according (7) depends
on the thickness of the layer of laminar convection and plasma
viscosity. We estimate the viscosity coefficient taking into account
the fact that the time of the generation of waves equals the length
of the cycle [14-16].

Taking into account that

$$ h \approx \lambda \approx \frac {l}{m}2
\pi R_I \cos \varphi \approx \frac{1,4\pi}{m} R_{Sun} \cos \varphi
$$

we can find the coefficient of viscosity:

$$ \nu \approx \frac {\pi^2 R_{Sun}^2 \cos^2\varphi}{m \Omega_{Sun}t_{g}^2 \sin \varphi}
\frac{sm^2}{sec} \eqno (8) $$

where $R_{Sun}$ is the radius of the Sun, m is is the order of
Rossby waves. For "active latitudes" the value of the coefficient of
viscosity can be compared with the value of the coefficient of
radiant viscosity $\nu_{\gamma}$ around the base of convective
envelope (the region of Rossby wave's generation):

$$\nu_{\gamma} \approx \frac {1}{3}\frac {c}{n \sigma_{T} } \approx (1.5 - 2) \cdot 10^{12} \frac {sm^2}{sec}  \eqno (9) $$

where $c$ is speed of light, , $\sigma_{T}$  - is the Thomson's
cross-section of scattering of photons by electrons and the
concentration of plasma n around the base of convective envelope is
equal to: $n\approx 5\cdot 10^{21}sm^{-3}$. It is a coincidence
speaks in favor of the above-described the physical nature of the
global magnetic activity. Approximately the same value of turbulent
viscosity is used in the  $\alpha\Omega$-dynamo model. The model of
multi-layered laminar convection using radiant viscosity (9) is
described in [15].

Rossby waves at different helio-latitudes and with different
wavelengths generate a very complex structure and evolution of the
poloidal field. Check of the physical picture of this poloidal field
formation according to the observations of the Sun only is
difficult. So we need the study of solar-type stars which are
analogs of the Sun of different stages of evolution [10, 11].

We estimate the relationship between the duration of the cycle of
activity and the effective temperature of the stars. The angular
velocity of rotation of the star, and its effective temperature are
related by the ratio $\Omega \sim T_{eff}~^4$, see Fig. 1.

In fully ionized hydrogen plasma its concentration depends on the
temperature, as $n \approx T^ {-\frac{3}{2}}$.

Using radiant viscosity (3) $\nu \approx \nu_\gamma \sim
T^{-\frac{3}{2}}\sim T_{eff}^{-\frac{3}{2}}$ and under assumption
that the time of generation of Rossby waves $t_g$ corresponds to
duration of the activity cycle $T_{cyc}$ from the formula (7) we
obtain the following connection between the star's duration of the
activity cycle and its effective temperature:

$$ T_{cyc} \approx t_R \sim T_{eff}~^{-\frac{5}{4}}  \eqno (10) $$

So we can see that relation (10) is consistent (within our
assumptions) with the dependencies which are showed at Fig. 2 and
Fig. 3.

Thus we show that theoretical dependence of the time of generation
of Rossby waves $t_g$ versus $T_{eff}$ (the basic parameter of a
star) describes well the connection between the star's duration of
the activity cycle $T_{cyc}$ (obtained from observations of
solar-type stars and the Sun, see Table 1) and their $T_{eff}$.

We have also to take into consider that solar cyclic activity is the
very important factor in learning of space weather behavior. Thus,
it becomes important to explain the nature of so-called "11-year"
cycles on the Sun and solar-type stars.

And so, for the quantitative analysis of the proposed physical
picture it is necessary to conduct the numerical experiments of the
magnetic hydrodynamics of a multilayer convective shell of the Sun.

The time interval $t_g$ which is necessary for the generation of
Rossby waves (see formula (1)) is equal to 2 years approximately
when we use the suitable values of the coefficient of viscosity (3).
But if of the coefficient of viscosity is ten times less then (3):
this value is not impossible, see [17] we get the period of the
Rossby waves generation is equal to  8 -10  years.

\section{Conclusions}

We offer the physical picture of the interrelationship of observed
properties of the local and global magnetic activity of the Sun. The
main new element in this picture is a hypothesis about the
possibility of the existence of at least two layers of the
convective shell of the solar-type stars. Around the ground level
the convective shell is the laminar convection layer, which consists
of giant convective cells. Thanks to the rotation on the surface of
this layer Rossby waves are formed. These waves have spiral
structure due to differential rotation. The formation of the laminar
convection is caused by a strong heating of plasma photons from the
zone of radiant heat transfer and high viscosity of radiant fully
ionized dense plasma. Viscosity of radiant plasma effectively
inhibits the directed motion of electrons more than directed motion
of protons. Therefore, in the Rossby waves the current appears, and
this current generates the poloidal field. This field is a primary
reason of the whole magnetic activity of stars. Above the layer of
the laminar convection the layer of turbulent convection extends. In
the layer of turbulent convection the primary poloidal field
generates a toroidal field of starts and process of generation of a
strong local magnetic activity solar-type stars in the medium and
equatorial latitudes starts. Local and global magnetic activities
are interconnected thanks to the existence of internal Rossby waves
and the primary poloidal field.

It is shown that in the framework of the hypothesis proposed in our
work about the existence of internal Rossby waves you can explain
the dependence of "the duration of the activity cycles - the
effective temperature" for 11-year and quasi-biennial cycles, see
Fig. 2, 3. The duration of the activity cycles according to the
order of magnitude is equal to the characteristic time of generation
of Rossby waves. Under the assumption that the time of generation of
Rossby waves $t_g$ corresponds to duration of the activity cycle
$T_{cyc}$ it was shown that empirical dependencies $T_{11}\sim
T_{eff}^{- 1,1}$ and $T_2\sim T_{eff}^{- 0.79}$ describes well the
dependencies between the star's duration of the activity cycle
$T_{cyc}$ and their $T_{eff}$ estimated by our model $ T_{cyc}
\approx t_R \sim T_{eff}~^{-\frac{5}{4}}$.

Our study is in good agreement with the fact of existence of the set
of periods of cyclical activity on the Sun. There exist the
cyclicities of activity with the next durations:  T11/2 - "5,5-year
cycle", T11/4 - quasi-biennial cycle, T11/8 - "1,3-year cycle" and
2T11 - "22-year", 4T11 - semi-century, 8T11 - century cyclicities.
Observations of this set of periods of cyclical activity (equal to
some parts of the main "11-year" period) show us that a wave nature
in the phenomenon of cyclicity of solar activity takes place. It
confirms our assumption that the duration of the activity cycles is
determined by the length of the Rossby waves.

Also it was shown that when we use the different estimations of the
value of the coefficient of viscosity at the bottom of convective
zone of the Sun, the time of generation of Rossby waves $t_g$ (which
corresponds to duration of the activity cycle$T_{cyc}$)   may have
the duration is approximately of 2 - 10 years.

Thus we emphasized the significant contribution of Rossby waves in
formation of magnetic cycles of stars and the Sun.

\bigskip
{\bf References}

1. Vitinsky YM, Kopecky M, Kuklin GB. The statistics of the spot
generating activity of the Sun, Moscow: Nauka; 1986.

2. Monin AS. The global hydrodynamics of the Sun. Physics-Uspekhi
Journal (UFN). 1980, V132(1), pp.123-166.

3. Babcock HW. The Topology of the Sun's Magnetic Field and the
22-Year Cycle. Astrophys. J. 1961, V133(2), pp.572-587.

4. Kollath Z, Olah K. Multiple and changing cycles of active stars.
Astron. and Astrophys.  2009, V501(2), pp.695-705.

5. Bruevich EA, Kononovich EV. Solar and Solar-type Stars
Atmosphere's Activity at 11-year and Quasi-biennial Time Scales.
Moscow University Physics Bulletin. 2011, V66(1), pp.72-77, arXiv:
1102.3976v1.

6. Rozgacheva IK, Bruevich EA. Model of laminar convection in
solar-type stars. Astronomical and Astrophysical Transactions.
2002;21(1):27-35, arXiv:1204.1148v1.

7. Parker EN. Cosmical Magnetic Fields. Oxford: Clarendon Press;
1979.

8. Mininni P, G?mez D, Mindlin B. A simple model of a stochastically
excited solar dynamo. Solar Phys. 2001, V201, pp.203-213.

9. Berdyugina SV, Moss D, Sokoloff D, Usoskin IG. Active longitudes,
nonaxisymmetric dynamos and phase mixing. Astron. and Astrophys.
2006, V445(2), pp.703-714.

10. Donahue RA, Saar SH, Baliunas SL. A Relationship between
Rotation Period in Lower Main-Sequence Stars and Its Observed Range.
Astrophys. J. 1996, V466(1), pp.384-391.

11. Baliunas SL, Donahue RA, Soon WH, Horne JH, Frazer J,
Woodard-Eklund L. et al. Chromospheric variations in main-sequence
stars. II. Astrophys. J. 1995, V438, pp.269-280.

12. Schnerr RS, Spruit HS. The brightness of magnetic field
concentrations in the quiet Sun. 2010;arXiv:1010.4792v3.

13. Seehafer N, Gellert M, Kuzanyan KM, Pipin VV. Helicity and the
solar dynamo, Adv. Space Res. 2003, V32(10), pp.1819-1833.

14. Gilman PA. A Rossby-wave dynamo for the Sun. I. Solar Physics.
1969, V8, pp.316-330.

15. Gilman PA. Solar rotation. Ann. Rev. Astron. and Astrophys.
1974, V12, pp.47-70.

16. Kichadinov LL. The differential rotation of the stars.
Physics-Uspekhi Journal (UFN). 2005, V175(5), pp.475-494.

17. Krivodubskij VN. The negative turbulent viscosity of solar
plasma. Izvestiya. KrAO. 2010, V106(1), pp.226-231.

\end{document}